\documentstyle[aps,multicol,amsmath]{revtex}

\begin{document}

\title{Langevin Equation for the Rayleigh model with finite-ranged interactions}
\author{Alexander V. Plyukhin and Jeremy Schofield,\\
{\small {\sl Chemical Physics Theory Group, Department of Chemistry,}}\\
{\small {\sl University of Toronto, Toronto, Ontario, Canada M5S 3H6}}}
\date{\today}
\maketitle

\begin{abstract}
Both linear and nonlinear Langevin equations are derived directly from
the Liouville equation for an exactly solvable model consisting of a Brownian particle 
of mass $M$ interacting with ideal gas molecules 
of mass $m$ via a quadratic repulsive potential.  Explicit microscopic
expressions for all kinetic coefficients appearing in these equations
are presented.
It is shown that the range of applicability of the Langevin equation,
as well as  statistical properties of random force,
may depend 
not only on the mass ratio $m/M$ but also by the parameter $Nm/M$, 
involving the 
average number $N$ of molecules in the interaction zone around the particle.
For the case of a short-ranged potential, when $N\ll 1$, analysis
of the Langevin equations yields previously obtained results for a
hard-wall potential in which only binary collisions are considered.
For the finite-ranged potential, when multiple collisions are important
($N\gg 1$), the model describes nontrivial dynamics on time 
scales that are on the order of the collision time, a regime that is
usually beyond the scope of
more phenomenological models.       
\end{abstract}


\section{Introduction}
The explicit derivation from first principles of the Langevin equation, 
describing the evolution of
a small number of variables in a complex system, 
is often necessary since in many cases the statistical properties of 
the random force, the range of applicability and even the form of the equation are far from 
evident~\cite{Zwanzig}. The need for microscopic considerations is especially compelling in the
case of a nonlinear Langevin equation where the usual phenomenological 
approach 
of adding stochastic terms to the deterministic nonlinear equation describing
the relaxation of targeted variables may be inadequate~\cite{VanKampen}. 
Although statistical properties of 
the random force in the nonlinear Langevin equation may be deduced 
phenomenologically in some cases~\cite{phenomen}, it is generally
necessary to start from the microscopic behavior of the system in
order to construct the appropriate form of the equation.    
 
The conventional systematic method for deriving the Langevin equation (LE) for a Brownian 
particle 
exploits Mori's projection-operator techniques~\cite{Zwanzig}, 
which allow the
transformation of the microscopic Liouville equation to a non-Markovian  
predecessor of the LE, generally known as the generalized Langevin
equation.  The LE can be then  obtained by a subsequent
perturbation expansion of the memory kernel appearing in the
generalized Langevin equation
using the square root ratio of the mass of a bath particle to that 
of the Brownian particle, $(m/M)^{1/2}$, as a
perturbation parameter $\lambda$.
While the first step in the derivation of the LE,
involving  rearrangement of the Liouville equation with projection operator
methods, is an exact algebraic procedure, the question of convergence
of the $\lambda$ expansion in the second step is subtle and can be strictly 
justified only under assumptions
which are difficult to prove in general~\cite{Ray1}.
Mazur and Oppenheim
developed an alternative projection-operator approach
more suitable to analyse 
the convergence of the $\lambda$ expansion~\cite{MO}.
The validity
of the perturbation analysis in both the Mori and the Mazur-Oppenheim 
approaches has been examined for an exactly solvable model consisting of
tagged particle motion in a harmonic
lattice~\cite{Deutch} . 
Unfortunately, the convergence properties of
the $\lambda$ series are trivial for this model, since all terms in the
$\lambda$ expansion higher than zeroth order vanish. As a result,
the memory kernel for the linear damping force in the LE does not
depend on the mass ratio, and all non-linear terms are identically 
zero~\cite{Ray2}.   

One goal of this paper is to present and analyze a simplified model 
that serves as a useful and non-trivial testing ground for examining
some of the subtle aspects of the theory of Brownian motion.  The
model considered here is a generalization of the well-known 
Rayleigh model of a Brownian particle of mass $M$ constrained to move in one 
dimension and 
subjected to collisions with an equilibrium ideal (non-interactive) gas of 
particles of mass $m$. The Rayleigh model is perhaps the oldest
model of non-equilibrium statistical physics~\cite{Rayleigh}, and has attracted
much attention over the years, with early work~\cite{Leb,VK,Hynes} focusing
on the model as a test of the systematic derivation of macroscopic kinetic 
equations for the heavy particle from the master equation.  More recent 
investigations  have examined the 
stationary and transient solutions of the asymmetric Rayleigh 
model in which the thermodynamic parameters characterizing the gas to
the left and right of the piston differ~\cite{piston}. 
In all these studies, the interaction between the Brownian particle
and the bath is assumed to be short-ranged with a negligible  collision 
time $\tau_c$. Only binary collisions are considered in this model because 
the range of interaction is assumed to be short
compared to the average distant between bath particles.
The usual starting point for analysis of the Rayleigh model with binary 
collisions is 
a Markovian master equation for particle's velocity distribution
function. The master equation is only an approximate form of
the fully microscopic Liouville equation, and valid only for time scales 
longer than $\tau_c$. 
To test many aspects of the theory of Brownian motion, it is essential to start from a fully
microscopic description of the dynamics directly from the Liouville
equation so that any non-Markovian character of bath correlations is
properly incorporated and particle dynamics on time scales less than
$\tau_c$ can be described. To this end, one may generalize the
interaction between the bath and Brownian particles from a hard-wall
to a parabolic repulsive potential.  The generalized model is
analytically solvable, allowing explicit calculation of all terms
appearing in the derivation of both the linear and nonlinear LE
beyond the binary collision approximation.  
We demonstrate that in addition to the small parameter $\lambda=(m/M)^{1/2}$,  
the character of the dynamics of a tagged particle can be governed 
by an additional  
parameter, $N^{1/2}\lambda$,  involving the average number $N$ 
of bath particles simultaneously interacting with the Brownian particle. 
For a large Brownian particle, it is shown that when $N\gg 1$ and
multiple collisions are important, the parameter of the formal
$\lambda$-expansion is actually $N^{1/2}\lambda$.

The paper is organized as follows:  In section 2, the Mazur-Oppenheim approach is reviewed to provide groundwork for 
all subsequent analysis. In section 3,  the structure of terms in the
$\lambda$ expansion is examined and presented in a convenient form. In section 4, 
the general formalism is applied to the Rayleigh model with 
a quadratic repulsive potential describing bath-Brownian particle
interactions, and the LE is derived for the heavy particle. 
The nonlinear LE is obtained in section 5 and various aspects of this
equation are discussed. Finally, a few concluding remarks are made in
section 6.

\section{Basic equations}
The Hamiltonian for a Brownian system composed of a tagged particle of mass $M$ in a bath of 
point particles of mass $m$ is
\begin{eqnarray}
H&=&\frac{P^2}{2M}+H_0,\\
H_0&=&\sum_i\frac{p_i^2}{2m}+U(x,X),
\end{eqnarray}
where $x=\{x_i\}$ and $p_i$ are positions and 
momenta of bath particles, $X$ and $P$ are those of the Brownian (or
tagged) particle, 
and $H_0$ is the Hamiltonian of the bath in the field of the tagged  
particle fixed at $X$.
One can expect that on average $P\sim\sqrt{M k_BT}$, where $k_B$ is Boltzmann's constant and $T$ is
the temperature, and that the scaled 
momentum $P_*=\lambda P$, where 
$\lambda=\sqrt{m/M}$, is of the same order as the average 
momentum of a bath particle.
In terms of scaled momentum, one can write the Liouville operator as
\begin{eqnarray}
{\mathcal L}&=&{\mathcal L}_0+\lambda {\mathcal L}_1,\\
{\mathcal L}_0&=&\sum_i\left\{\frac{p_i}{m}\frac{\partial}{\partial x_i}+
F_i\frac{\partial}{\partial p_i}\right\},\\
{\mathcal L}_1&=&\frac{P_*}{m}\frac{\partial}{\partial X}+
F\frac{\partial}{\partial P_*},
\end{eqnarray}
where $F_i=-\nabla_{x_i}U$ and $F=-\nabla_XU$ are the forces on the {\it i}-th bath particle
and on the Brownian particle, respectively. The operator ${\mathcal L}_0$ dictates the dynamics 
of the bath in the field of the fixed Brownian particle.

If the mass of the tagged particle is large (i.e. a Brownian particle), one 
might intuitively expect that
inertial effects of the particle's motion are small and that
the force on the particle, $F(t)=e^{{\mathcal L}t}F$,  is close to the 
pressure 
force, i.e. to the force on the fixed tagged particle, 
\begin{equation}
F_0(t)\equiv e^{{\mathcal L}_0t}F.
\end{equation}
In the Mazur-Oppenheim approach~\cite{MO}, the force $F(t)$ is
decomposed using the projection operator $\mathcal P$ 
which averages a dynamical variable $A$ over the canonical distribution
$\rho=Z^{-1}exp(-\beta H_0)$, for bath variables at fixed position of the tagged particle, 
\begin{equation}
{\mathcal P}A=\langle A\rangle \equiv\int \rho A \prod_i dx_i\, dp_i,
\end{equation}
where $Z$ is the canonical partition
function and $\beta = 1/k_{B}T$.
Using the operator identity~\cite{identity}
\begin{equation}
e^{{\mathcal(A+B)}t}=e^{{\mathcal A}t}+\int_0^t d\tau 
e^{{\mathcal A}(t-\tau)}{\mathcal B} e^{({\mathcal A+B})\tau},
\label{identity}
\end{equation}
with ${\mathcal A}={\mathcal L}$ and ${\mathcal B}=-{\mathcal P}{\mathcal L}$, 
one may formally decompose the force $F(t)$ on the tagged particle
into a ``random'' part and a remainder as
\begin{eqnarray}
F(t)&=&F^\dagger(t) +\int_0^t d\tau\,\,
e^{{\mathcal L}(t-\tau)}{\mathcal P}{\mathcal L}F^\dagger(\tau),
\label{randomForce}
\end{eqnarray}
where $F^\dagger(t)=e^{{\mathcal Q}{\mathcal L}t}F$ and 
${\mathcal Q}=1-{\mathcal P}$. The factor 
${\mathcal P}{\mathcal L}F^\dagger(\tau)$ in the integral in Eq.~(\ref{randomForce})
can be simplified taking into account the orthogonality of ${\mathcal P}$
and ${\mathcal L}_0$ (i.e. ${\mathcal P}{\mathcal L}_0=0$), and the equality 
\begin{equation}
\langle\nabla_X F^\dagger(t)\rangle=-\beta\langle F_0
F^\dagger(t)\rangle ,
\label{gradient}
\end{equation}
which can be derived by integration by parts.
As a result, one obtains the following 
exact equation of motion for the scaled momentum of the tagged particle,
\begin{equation}
\frac{dP_*(t)}{dt}=\lambda F^\dagger(t) +
{\lambda}^2\int_0^t d\tau\,\,
e^{{\mathcal L}(t-\tau)}\Bigl(\nabla_{P_*}-\frac{\beta}{m}P_*\Bigr)
\langle FF^\dagger(\tau)\rangle, 
\label{EM}
\end{equation}
where $F^{\dagger}(t)$ is 
a zero-centered random force obeying
$\langle F^{\dagger}(t)\rangle={\mathcal P}e^{{\mathcal Q}{\mathcal L}t}F=0$.

The random force $F^{\dagger}(t)=e^{({\mathcal L}_0+\lambda{\mathcal
Q}{\mathcal L}_1)t}F$ can be further expanded in terms of the mass
ratio parameter $\lambda$ using identity (\ref{identity}) to obtain
\begin{eqnarray}
F^{\dagger}(t) &=& F_0(t)+
\lambda\int_0^t dt_1\, e^{{\mathcal L}_0(t-t_1)}{\mathcal Q}{\mathcal L}_1 F_0(t_1)+ \\
&&\lambda^2\int_0^t dt_1 \int_0^{t_1} dt_2 \,
e^{{\mathcal L}_0(t-t_1)}{\mathcal Q}{\mathcal L}_1 
e^{{\mathcal L}_0(t_1-t_2)}{\mathcal Q}{\mathcal L}_1 F_0(t_2)+\cdots\nonumber
\end{eqnarray}
Similarly, the kernel $K(t) = \langle FF^{\dagger}(t) \rangle$
appearing in the exact equation of motion (\ref{EM}) of the tagged
particle may be 
expanded in a power series in $\lambda$
\begin{eqnarray}
K(t)&=&\langle FF^{\dagger}(t)\rangle=
\sum_l K_l(t), \label{expand} \\
K_0(t)&=&\lambda^0\langle FF_0(t)\rangle,\nonumber\\
K_1(t)&=&\lambda^1\int_0^t dt_1\, C_1(t,t_1), \nonumber \\
K_2(t)&=&\lambda^2\int_0^t dt_1 \int_0^{t_1} dt_2 \,
C_2(t,t_1,t_2) \cdots\nonumber
\end{eqnarray}
where the correlation functions $C_l$ are defined to be
\begin{eqnarray}
C_1(t_0,t_1) &=&
\left\langle F \left(e^{{\mathcal L}_0(t_0-t_1)}{\mathcal Q}{\mathcal L}_1\right) F_0(t_1)
\right\rangle,\label{CF}\\
C_2(t_0,t_1,t_2) &=& \left\langle F 
\left(e^{{\mathcal L}_0(t_0-t_1)}{\mathcal Q}{\mathcal L}_1\right)
\left(e^{{\mathcal L}_0(t_1-t_2)}{\mathcal Q}{\mathcal L}_1\right)
F_0(t_2)\right\rangle, \nonumber\\
C_l(t_0,t_1,\cdots,t_{l}) &=&
\left\langle F \Bigl(\prod_{i=1}^{l} e^{{\mathcal L}_0(t_{i-1}-t_i)}{\mathcal Q}{\mathcal L}_1\Bigr) 
F_0(t_{l})\right\rangle\nonumber.
\end{eqnarray}
The truncation of the $\lambda$-expansion to zeroth order,
$K(t)\approx K_0(t)$, leads from Eq.(\ref{EM}) 
directly to the generalized Langevin equation
\begin{equation}
\frac{dP_*(t)}{dt}=\lambda F^{\dagger}(t) -
{\lambda}^2\int_0^t d\tau\,\,
M_0(\tau) P_*(t-\tau),
\label{GLE}
\end{equation}
where 
\begin{equation}
M_{0}(t) = \frac{\beta}{m}K_{0}(t) = \frac{\beta}{M}\langle F F_{0}(t)
\rangle .
\label{M0}
\end{equation}
This approximation is sensible provided the correlation functions
$C_{l}$ appearing at higher order in the $\lambda$-expansion decay on a
similar $\lambda$-independent time scale $\tau_c$
characteristic of motions of the fixed-particle system 
(i.e. governed by ${\mathcal L}_0$). 
Mazur and Oppenheim~\cite{MO} 
succeeded in proving that this is the case 
assuming the factorization properties
$
\langle A(t_1)e^{{\mathcal L}_0t}B(t_2)\rangle
\stackrel{t>\tau_c}{\longrightarrow}
\langle A(t_1)\rangle\langle B(t_2)\rangle .
$

While being formally non-local in time, Eq.~(\ref{GLE}) can actually be 
written in a form that is local in time by expanding $P_{*}(t - \tau
)$ around $\tau=0$ to obtain 
\begin{equation}
\frac{dP_*(t)}{dt}=\lambda F^{\dagger}(t) -
{\lambda}^2\gamma_0(t)P_*(t), 
\label{TDLE}
\end{equation}
where $\gamma_0(t)=\int_0^t d\tau M_0(\tau)$. 
The non-local correction terms to this approximation are of the form~\cite{comment}
\begin{eqnarray}
\lambda^2\int_0^t d\tau M_0(\tau)
\int_{t-\tau}^t d\tau'\,\dot{P}_*(\tau')\sim \lambda^3 ,
\label{estim1}
\end{eqnarray}
and are therefore of higher order in $\lambda$.
Naturally, this analysis is pertinent only if 
the characteristic time $\tau_c$ for
decay of $M_0(t)$ does not depend on $\lambda$. 

The local equation (\ref{TDLE}) is applicable on arbitrary time scales 
and for $t\gg \tau_c$ assumes
the form of the conventional
LE  with a time-independent damping coefficient $\gamma_0=\int_0^\infty dt
M_0(t)$. It is then evident from (\ref{TDLE}) that the autocorrelation
function of the momentum of a
heavy particle decays on a time-scale $\tau_p\sim \lambda^{-2}$
that is much longer than the characteristic time of the bath
$\tau_c\sim\lambda^0$.  One can therefore expect the local form of the
LE with time-independent damping to be a good approximation for Eq.~(\ref{TDLE})
except at short times determined by $t<\tau_c$. 
  
It will be shown below that for homogeneous bath 
$K_1(t)=0$, so the next approximation 
for the $\lambda$-expansion (\ref{expand}) is of the form 
$K(t)\approx K_0(t)+K_2(t)$. The equation of motion for the momentum
of the tagged particle in this case includes a non-linear damping term
of third order in $P_*$. 
This equation will be considered in section 5.

\section{Structure of the terms in $\lambda$-expansion}
To examine the convergence properties of the $\lambda$-expansion and
other features of the projection operator derivation of the LE, 
the structure of the correlation 
functions $C_l$ defined in (\ref{CF}) which appear in the expansion of
the memory function must be analyzed.
Although only the functions $C_0$ and $C_2$ are needed to obtain the 
nonlinear LE to lowest order in $\lambda$, it is useful  to 
know general properties of $C_l$.  

By inspection of the symmetry properties of the system, it is
immediately apparent that the correlation functions $C_{2n+1}$
corresponding
to odd powers of $\lambda$ contain an odd number of factors $F$ and 
$\nabla_X$, and therefore  vanish for isotropic systems. In fact, 
in the absence of external field, the dependence on the particle coordinate
appears only through the difference $x_i-X$, and it is useful to
introduce new variables $q_i=x_i-X$. Since 
the vectors $F=-\nabla_X U=\sum_i\nabla_{q_i} U$ 
and $\nabla_X=-\sum_i\nabla_{q_i}$ have negative parity and the Hamiltonian 
$H_0$ is invariant  with respect to
transformation $\{q_i\to-q_i,\,\,p_i\to-p_i\}$, the correlation functions
$C_{2n+1}(t,t_1,...t_{2n+1})$ vanish.

Correlation functions of even orders $C_{2n}$ do not vanish and 
have a rather complicated structure. 
For notational simplicity, we restrict the analysis to the case of 
one-dimensional diffusion, expecting no physical features in higher 
dimensions. For future development, it is convenient to define
\begin{eqnarray}
G_0(t) &=& F_0(t),\nonumber\\
G_1(t,t_1) &=& S(t-t_1)F_0(t_1),\nonumber\\
G_2(t,t_1,t_2) &=&
S(t-t_1)S(t_1-t_2)F_0(t_2),\nonumber\\
G_s(t,t_1,\cdots,t_s) &=&
S(t-t_1)\cdots S(t_{s-1}-t_s)F_0(t_s),
\label{Gs}
\end{eqnarray}
where
\begin{equation}
S(t_i-t_l)=e^{{\mathcal L}_0(t_i-t_l)}\frac{\partial}{\partial X}.
\end{equation}
Note the property
\begin{eqnarray}
\langle G_{i_1}G_{i_2}\cdots G_{i_l}\rangle=0
\label{sym}
\end{eqnarray}
 which holds for arbitrary time arguments
when $i_1+i_2+\cdots +i_l$ and $l$ have different parities.
For example, $\langle G_i\rangle$ and 
$\langle G_0G_0G_i\rangle$ vanish for even $i$, while
$\langle G_0G_i\rangle$ and $\langle G_0G_0G_0G_i\rangle$ are zero
for odd $i$.

Using the definitions above and according to Eqs.~(\ref{CF}),
the second order correlation function $C_2$ can be 
written as 
\begin{eqnarray}
C_2(t,t_1,t_2) &=& \left(\frac{P_*}{m}\right)^2
\langle G_0G_2(t,t_1,t_2)\rangle \nonumber\\
&&+ \frac{1}{m}\langle G_0G_0(t-t_1)G_1(t,t_2)\rangle \nonumber\\
&&-\frac{1}{m}\langle G_0G_0(t-t_1)
\rangle\langle G_1(t,t_2)\rangle.
\end{eqnarray}
Using cumulants, denoted by $\langle\!\langle\ \cdots \rangle\!\rangle$ 
and defined via the relations,
\begin{eqnarray} 
\langle A\rangle &=&\langle\!\langle A_1\rangle\!\rangle,\nonumber\\
\langle A_1A_2\rangle &=& \langle A_1\rangle\langle A_2\rangle+
\langle\!\langle A_1A_2\rangle\!\rangle,\nonumber\\
\langle A_1A_2A_3\rangle &=& \langle A_1\rangle\langle A_2\rangle
\langle A_3\rangle+\nonumber\\
&&\langle A_1\rangle \langle\!\langle A_2A_3\rangle\!\rangle+
\langle A_2\rangle \langle\!\langle A_3A_1\rangle\!\rangle+
\langle A_3\rangle \langle\!\langle A_2A_1\rangle\!\rangle+\nonumber\\
&&\langle\!\langle A_1A_2A_3\rangle\!\rangle, \nonumber\\
\vdots \nonumber
\label{cumexp}
\end{eqnarray}
the function $C_2$ can be written as 
\begin{eqnarray}
C_2(t,t_1,t_2) &=&
\left(\frac{P_*}{m}\right)^2\langle\!\langle 
G_0G_2(t,t_1,t_2)\rangle\!\rangle+\nonumber\\
&&\frac{1}{m}
\langle\!\langle G_0G_0(t-t_1)G_1(t,t_2)\rangle\!\rangle.
\label{C2}
\end{eqnarray}
Note that the zeroth order kernel $K_0$ is also a cumulant,
$K_0=\langle FF_0(t)\rangle=
\langle\!\langle G_0G_0(t)\rangle\!\rangle$. 
The relevance of cumulant representation for $C_l$ follows from 
the fact that one expects the cumulants to have similar scaling
properties with respect to parameters of the system independent of
their order.  For example, it will be established in the next section that cumulants 
$\langle\!\langle G_{i_1}G_{i_2}\cdots G_{i_l}\rangle\!\rangle$
of any order are linear functions of the average number $N$ of particles in the interaction zone around
the particle for the Rayleigh model. 
Therefore, for this model, 
the first two non-vanishing terms 
in the $\lambda$-expansion (\ref{expand}), namely $K_0$ and $K_2$, 
are both linear in $N$. 

Consider the correlation functions $C_{2l}$ with $l\ge 1$.  It is
tedious, though not difficult to establish that these correlation functions 
are of order $l$ in cumulants 
$\langle\!\langle G_{i_1}G_{i_2}\cdots G_{i_l}\rangle\!\rangle$. 
Note that $C_{2l}$ has the contributions which are
of order $l+2$ or less in $G_i$. First consider
the terms of maximal order $l+2$:  One type of these terms are
the correlation functions of the form 
\begin{eqnarray}
\langle G_0G_0G_{i_1}G_{i_2}\cdots G_{i_l}\rangle,
\label{structure}
\end{eqnarray}
where here and below time arguments  have been omitted for brevity.
The indices $i_1,\cdots,i_l$ may take any values from 
the set $\{ 0,1,\cdots,l \}$ provided 
\begin{equation}
i_1+i_2+\cdots+i_l=l.
\label{condition}
\end{equation}
For example, for $l=1$ the correlation function of maximal order in $G_i$ 
which contributes to $C_2$
is $\langle G_0G_0G_1\rangle$, as can be seen from Eq.~(\ref{C2}).  For $l=2$  the function $C_4$ includes 
contributions of the form $\langle G_0G_0G_0G_2\rangle$ and
$\langle G_0G_0G_1G_1\rangle$, and so on.

All other contributions of order $l+2$ in $G_i$ can be written as 
products of correlation functions of lower orders which
can be obtained dividing the sequences
\begin{equation} 
G_0(\tau_1)G_0(\tau_2)G_{i_1}(\tau_3)G_{i_2}(\tau_4)\cdots G_{i_l}(\tau_{l+2})
\end{equation}
into groups in all possible ways without permuting functions.
For example, for $l=2$ the three remaining terms are
\begin{eqnarray}
&&\langle G_0(\tau_1)G_0(\tau_2)\rangle\langle G_0(\tau_3)G_2(\tau_4)
\rangle,\nonumber\\
&&\langle G_0(\tau_1)G_0(\tau_2)\rangle\langle G_1(\tau_3)G_1(\tau_4)
\rangle,\nonumber\\ 
&&\langle G_0(\tau_1)G_0(\tau_2)G_1(\tau_3)\rangle\langle G_1(\tau_4)
\rangle,\nonumber\\
&&\langle G_0(\tau_1)G_0(\tau_2)\rangle\langle G_1(\tau_3)\rangle\langle 
G_1(\tau_4)\rangle\nonumber.
\end{eqnarray}
Clearly, the terms of maximal order in cumulants corresponds to
the case when all indices $i_s$ in Eqs.~(\ref{structure}) and 
(\ref{condition}) are equal to one. Then
the cumulant expansion of all such terms contains the contribution 
\begin{eqnarray}
\langle G_0G_0\rangle 
\langle G_1\rangle
\langle G_1\rangle  
\cdots
\langle G_1\rangle, 
\label{max}
\end{eqnarray} 
which contains $l$ factors  $\langle G_1(t_i)\rangle$ and, therefore, 
is of order $l+1$ in cumulants.
However, all terms of order $l+2$ in $G_i$ described above 
enter in $C_{2l}$ via combinations
\begin{eqnarray}
&&\langle G_0G_0G_{i_1}G_{i_2}\cdots G_{i_l}\rangle-
\langle G_0G_0\rangle\langle G_{i_1}G_{i_2}\cdots G_{i_l}\rangle,\nonumber\\
&&\langle G_0G_0G_{i_1}\rangle\langle G_{i_2}\cdots G_{i_l}\rangle-
\langle G_0G_0\rangle\langle G_{i_1}\rangle\langle 
G_{i_2}\cdots G_{i_l}\rangle, \nonumber\\
&&\langle G_0G_0G_{i_1}G_{i_2}\rangle\langle G_{i_3}\cdots G_{i_l}\rangle-
\langle G_0G_0\rangle\langle G_{i_1}G_{i_2}\rangle
\langle G_{i_3}\cdots G_{i_l}\rangle,
\cdots\nonumber
\end{eqnarray}
which is a consequence of a presence of the operator 
${\mathcal Q}$ (first from the left) in the definition
of $C_i$ (see Eq. (\ref{CF})).
As a result, the contributions (\ref{max}) involving $l+1$ cumulant
factors cancel,
and the maximal order in cumulants of surviving terms in the
expression for $C_{2l}$ contain at most $l$ cumulant factors.

Having established that the contributions to $C_{2l}$ from terms of maximal 
order $l+2$ in $G_i$ are of order $l$ in cumulants, it is clear that
the terms of order $l+1$ and lower in $G_i$ cannot 
contain more that $l$ cumulant factors since such terms
contain at least one isolated factor of $G_0$ whose
average vanishes. Hence $C_{2l}$ contains {\it at most} $l$
cumulant factors.  

In the next section, we examine the consequences of the cumulant
expansions of the memory function for a specific system, namely, the
Rayleigh model with a repulsive parabolic potential.
It will be demonstrated that for the Rayleigh model, cumulants are linear
functions of the average number 
$N$ of particles in the interaction zone around the particle.  This,
in turn, implies that the terms in the $\lambda$-expansion behave as
\begin{equation}  
K_{2l}\sim (N+N^2+\cdots+N^l)\lambda^{2l}
\label{exp}
\end{equation}
for $l> 1$.  Clearly, for a large particle and/or long-ranged
potential leading to $N\gg 1$, these results suggest that $K_{2l}\sim
N^l\lambda^{2l}$,  demonstrating that
the actual parameter of $\lambda$-expansion for this exactly solvable
model is in fact $N^{1/2}\lambda$. 
On the other hand, for a short-ranged potential, when $N\ll 1$, 
all terms in the expansion are linear in $N$,  and one sees that $K_{2l}\sim N\lambda^{2l}$, 
and the effective small parameter of $\lambda$-expansion is in fact
the square-root of the mass ratio. To further illustrate this analysis, the
explicit form of the cumulant expansion of $C_{4}$ is presented in
Appendix A.

\section{The linear Langevin Equation for a heavy particle or idealized piston}
Consider the random motion of a piston of mass $M$ and cross-sectional area $S$
subjected to  collisions with an ideal gas
particles of mass $m$. The gas particles and the piston are constrained to 
move in one dimension perpendicular to the piston faces. 
The velocity distribution of incident particles $f_M(v)$
before collision with the piston
is Maxwellian with inverse temperature $\beta$, namely
\begin{equation}
f_M(v)=\left(\frac{m\beta}{2\pi}\right)^{1/2}
exp\left(-\frac{1}{2}\beta mv^2\right).
\end{equation}   
The piston-particle interaction is assumed to be described by a purely 
repulsive  
parabolic potential. For particles to the left of the piston the
interaction potential between a gas particle and the piston is 
\begin{equation}
U_l = 
\begin{cases}
\frac{1}{2}k_f\left(x-X_l \right)^{2} & x>X_l, \\
0 & x<X_l,
\end{cases}
\end{equation}
where $k_f$ is a force constant, $x$ is the coordinate of the gas particle,
$X_l=X_{lf}-a$ the boundary of the piston-particle interaction zone,
$X_{lf}$ is the coordinate of the left face of the piston, and
$a$ is the width of the interaction zone.  Similarly, the gas
particle-piston potential for the particles to the right
of the piston has the analogous form
\begin{equation}
U_r = 
\begin{cases}
\frac{1}{2}k_f \left(x-X_r \right)^{2} & x<X_r, \\
0 & x>X_r,
\end{cases}
\end{equation}
where $X_r=X_{rf}+a$, and $X_{rf}$ is the position of the piston's
right face. 
We assume that the temperature is low enough 
(or $k_f$ is sufficiently large)
so that the probability
for a particle to reach the piston's surface is negligible.

In the previous sections it was established that the dynamics of the piston
can be deduced from time correlation functions describing 
molecular motion in the field of the piston fixed in space.
Below we derive the explicit expressions for these correlation 
functions in the thermodynamic limit,  
neglecting recollisions of the piston and gas particles
due to the finite size of the bath.

Consider the force on the left side of the fixed piston,
\begin{equation}
F_l(t)= -k_f \sum_i q_i(t)\theta\Bigl( q_i(t) \Bigr),
\end{equation}
where $q_i=x_i-X_l$ is the position of gas particle $i$ relative to
the boundary of the interaction zone, 
$\theta (x)$ is the 
step-function, and summation over index $i$ is over all particle in the 
tube of diameter $S$ to the left of the piston.
In this section we omit for brevity the subscript
 $0$ for the force on the fixed piston.
The simplifying feature of parabolic potential is that the time $\tau_c$
that a particle spends in the interaction zone of the fixed piston is independent of 
the initial velocity of the particle and given by $\tau_c=\pi/\omega$, where 
$\omega=\sqrt{k_f/m}$. 
At a given time $t$, the only gas particles in the interaction zone
are those that had positive velocities and coordinates $q$ in the
interval $-v\tau_c<q<0$ at time $t-\tau_c$.
At time $t$, the position of the gas particle is determined by
$q(t)=(v/\omega)\sin\omega(t-t_{in})$, where $t_{in}=t-\tau_c-q/v$
corresponds to the time at which the gas particle enters the
interaction region and $q$ is the 
position of the gas particle at time $t-\tau_c$.
It then follows that 
$q(t)=(v/\omega)\sin\omega (\tau_c+q/v)=-(v/\omega)\sin\omega q/v$,
which implies that the total instantaneous force on the left side of the fixed piston at time $t>0$ 
can be written as
\begin{equation}
F_l(t)=-k_f \int\limits_0^\infty dv\int\limits_{-v\tau_c}^{0}dq \,\,
N(X_l+q,v;t-\tau_c )\,\,\,\frac{v}{\omega}\sin \frac{\omega q}{v}.
\label{force1}
\end{equation}
In Eq.~(\ref{force1}), $q=x-X_l$ and $N(x,v;t)$ is the microscopic
linear density of particles defined by
\begin{equation}
N(x,v;t)=\sum_{i}\delta\Bigl(x-x_i(t)\Bigr)\,\delta\Bigl(v-v_i(t)\Bigr).
\end{equation}
Similarly, the total instantaneous force acting on the right side of the piston is
\begin{equation}
F_r(t)=-k_f \int\limits_{-\infty}^0 dv\int\limits_0^{-v\tau_c}dq \,\,
N(X_r+q,v;t-\tau_c )\,\,\,\frac{v}{\omega}\sin\frac{\omega q}{v},
\label{force11}
\end{equation}
where $q=x-X_r$.

For a particle outside the interaction zone of the fixed piston (i.e. for 
$x<X_l$, and $x>X_r$), 
the average linear density of particles 
is 
$\langle N(x,v)\rangle=nSf_M(v)$, where $n$ is the
total (three-dimensional) density of bath particles, and $S$ is the
cross-sectional area of the piston.
It then follows from  Eqs.~(\ref{force1}) and (\ref{force11})
that the average force acting on
the left $\langle F_{l} \rangle$ and the right $\langle F_r \rangle$ sides of the piston are 
$\langle F_l\rangle=-\langle F_r\rangle=nS/\beta$.

It is straightforward to show that 
the stationary distribution in the vicinity of the fixed piston, 
including
the interaction zone, assumes Boltzmann's form
\begin{equation}
\langle N(x,v)\rangle=nSf_M(v)\exp(-U(x)/k_BT).  
\end{equation}
To calculate the force correlation functions required to analyze the
damping terms in the LE, 
correlation functions of the form
$\langle N(Y_1)N(Y_2)\cdots N(Y_s)\rangle$ must be evaluated, where
$Y$ denotes the position-velocity pair $(x,v)$. It is sufficient 
to consider only the case when time arguments 
are equal for all functions, since time displacement can be
transformed into spatial displacement for a free particle, i.e.
\begin{equation}
N(x,v;t+t_1)=N(x-vt_1,v;t).
\label{property}
\end{equation}
Note that the product $N(Y_1)N(Y_2)$ can be written as 
\begin{eqnarray}
N(Y_1)N(Y_2) &=& \sum_{i,j}\delta (Y_1-Y_i)\delta (Y_2-Y_j) \nonumber \\
&=&\sum_{i}\delta (Y_1-Y_i)\delta (Y_2-Y_i)+
\sum_{i\ne j}\delta (Y_1-Y_i)\delta (Y_2-Y_j). \nonumber
\end{eqnarray}
Since $\delta (Y_1-Y_i)\delta (Y_2-Y_i)=
\delta (Y_1-Y_2)\delta (Y_1-Y_i)$, for the ideal gas system one obtains
\begin{eqnarray}
\langle N(Y_1)N(Y_2)\rangle=\delta (Y_1-Y_2)\langle N(Y_1)\rangle+
\langle N(Y_1)\rangle\langle N(Y_2)\rangle.
\label{dencor1}
\end{eqnarray}
For the three-point correlation function, the same arguments lead to
the result
\begin{eqnarray}
\langle N(Y_1)N(Y_2)N(Y_3)\rangle &=&
\delta (Y_1-Y_2)\delta(Y_2-Y_3)\langle N(Y_1)\rangle\label{dencor2}\\
&+&\delta (Y_1-Y_2)\langle N(Y_1)\rangle\langle N(Y_3)\rangle+
\delta (Y_1-Y_3)\langle N(Y_1)\rangle\langle N(Y_2)\rangle\nonumber\\
&+&\delta (Y_2-Y_3)\langle N(Y_1)\rangle\langle N(Y_2)\rangle+
\langle N(Y_1)\rangle\langle N(Y_2)\rangle\langle N(Y_3)\rangle.
\nonumber
\end{eqnarray}
Eqs.~(\ref{dencor1}) and (\ref{dencor2}) are the cumulant 
expansions of $\langle N(Y_1) N(Y_2)\rangle$ and $\langle N(Y_1)
N(Y_2) N(Y_3) \rangle$, where the cumulants
\begin{eqnarray}
\langle\!\langle N(Y_1)N(Y_2)\rangle\!\rangle &=&
\delta (Y_1-Y_2)\langle N(Y_1)\rangle,\nonumber\\
\langle\!\langle N(Y_1)N(Y_2)N(Y_3)\rangle\!\rangle &=&
\delta (Y_1-Y_2)\delta(Y_2-Y_3)\langle N(Y_1)\rangle,\\
\langle\!\langle N(Y_1)N(Y_2)\cdots N(Y_s)\rangle\!\rangle &=&
\delta (Y_1-Y_2)\delta(Y_2-Y_3)\cdots\delta(Y_{s-1}-Y_s)\langle N(Y_1)\rangle
\nonumber
\label{cumdens}
\end{eqnarray}
are proportional to the equilibrium density $n$ of gas particles.

To order $\lambda^2$, the dynamics of the piston is described 
by the LE (\ref{TDLE}) with a time-dependent damping coefficient,
$\gamma_0(t)=\frac{\beta}{m}\int_0^t d\tau\langle FF_0(\tau)\rangle$, 
where the evolution of 
the total force $F_0=F_l+F_r$ is determined by the constrained
piston-bath Liouville operator ${\mathcal L}_0$.  Since $\langle F_l\rangle =-\langle F_r\rangle$ and 
$\langle F_lF_l(t)\rangle=\langle F_rF_r(t)\rangle$, 
\begin{equation}
\langle FF_0(t)\rangle=2\langle F_lF_l(t)\rangle-2\langle F_l\rangle^2=
2\langle\!\langle F_lF_l(t)\rangle\!\rangle.
\label{Kzero}
\end{equation}
From Eqs.~(\ref{Kzero}) and (\ref{force1}), one sees that 
$\langle FF_0(t)\rangle$ can be expressed in terms of the cumulant 
$\langle\!\langle N(x,v;t-\tau_c)N(x',v';-\tau_c)\rangle\!\rangle$. 
Using property (\ref{property}) and Eq. (\ref{cumdens}),
the cumulant may be rewritten as 
\begin{eqnarray}
\langle\!\langle N(x-vt,v;-\tau_c)N(x',v';-\tau_c)\rangle\!\rangle=
\delta(x-vt-x')\delta(v-v')nSf_M(v).
\label{twopoint}
\end{eqnarray}
Then, using Eqs.~(\ref{Kzero}), (\ref{force1}) and
(\ref{twopoint}), one obtains
\begin{eqnarray}
\langle FF_0(t)\rangle=\frac{nSk_{f}^2}{\omega^3}\theta(\tau_c-t)\Bigl\{
\sin\omega t+\omega(\tau_c-t)\cos\omega t\Bigr\}
\int_0^\infty \,dv f_M(v)v^3.
\label{K00}
\end{eqnarray}
This can be
re-expressed in the compact form
\begin{equation} 
\langle FF_0(t)\rangle
=N\omega^2\,\frac{m}{\beta}\,\xi_0(t),
\label{K0}
\end{equation}
where $\xi_0(t)$ is a dimensionless function
(see Fig. 1) given by
\begin{eqnarray}
\xi_0(t)=\sqrt{\frac{2}{\pi}}\,\theta(\tau_c-t)\Bigl\{ \sin\omega t+
(\pi-\omega t)\cos\omega t\Bigr\},
\label{xi0}
\end{eqnarray}
and 
\begin{equation}
N=nS\frac{\langle v^2\rangle^{\frac{1}{2}}}{\omega}
\label{alpha}
\end{equation}
is the average number of particles in the shell of thickness 
$l=\sqrt{\langle v^2\rangle}/\omega$ around the piston.
The parameter $l$  specifies the length at which the average bath particle penetrates the interaction zone, and $N=nSl$
is the average number of particles in the layer of thickness $l$ around
the piston. Thus, it is evident that in addition to the mass ratio 
$\lambda^2=m/M$, 
the relevant physics depends strongly on another characteristic
parameter of the system, namely,
$N\lambda^2$,  which can be interpreted as 
the ratio of total mass $M_{*}=mnSl$ of bath particles
in the layer of thickness $l$ in the vicinity of the piston to the
mass $M$ of the piston.

With these definitions in hand, the time dependent damping coefficient
in the LE (\ref{TDLE}) can be written as
\begin{eqnarray}
\gamma_0(t)=\frac{\beta}{m}\int_0^t d\tau\langle FF_0(t)\rangle=
\omega N\zeta_0(t),
\end{eqnarray}
and therefore
the LE~(\ref{TDLE})
assumes the form
\begin{eqnarray}
\frac{dP_*(t)}{dt}=\lambda F^\dagger(t) -
\omega\,{\lambda}^2 N \,\zeta_0(t)\, P_*(t),  
\label{gle1}
\end{eqnarray}
where the damping  function $\zeta_0(t)=\omega\int_0^td\tau \xi_0(\tau)$  
is given by
\begin{eqnarray}
\zeta_0(t)=\sqrt{\frac{2}{\pi}}\theta(\tau_c-t)
\Bigl\{2(1-\cos\omega t)+(\pi-\omega t)\sin\omega t\Bigr\}+
4\sqrt{\frac{2}{\pi}}\theta(t-\tau_c)
\label{zeta0}
\end{eqnarray}
(see Fig.1). This expression describes the interesting
time-development of the dissipative force, an aspect of the dynamics
that is outside the scope of more phenomenological
models. For $t<\tau_c$, Eq.~(\ref{gle1}) describes essentially 
non-exponential relaxation of the momentum 
$\langle P(t)\rangle=P(0)e^{-\chi(t)}$ with 
\begin{eqnarray}
\chi(t)=\lambda^2N\omega\int_0^td\tau\zeta_0(\tau)=
\lambda^2N\sqrt{\frac{2}{\pi}}\Bigl\{\pi+2\omega t-(\pi-\omega t)
\cos\omega t-3\sin\omega t\Bigr\}.
\end{eqnarray}
For small $\omega t$, $\chi(t)\sim (\omega t)^2$.
On a time-scale $t>\tau_c$ the damping function reaches its plateau value
and the Markovian limit of the Langevin equation, in which the
damping coefficient is independent of time, is recovered:
\begin{eqnarray}
\frac{dP_*(t)}{dt}=\lambda F^\dagger(t) -
4\sqrt{\frac{2}{\pi}}\, \omega\,{\lambda}^2N\, P_*(t).
\label{normal_LE}
\end{eqnarray}
Note that the characteristic time for relaxation of the momentum 
$\tau_p=\omega^{-1}(\lambda^2N)^{-1}$ is governed by the parameter $\lambda^2N$,
rather than $\lambda^2$.  It is also interesting to observe that the
average number of collisions of bath particles with the piston for $t
\sim \tau_p$ is of order $nS\langle v^2\rangle^{1/2}\tau_p=\lambda^{-2}$
and depends neither on temperature, nor on the piston's size, but only 
on the mass ratio.

Equation (\ref{normal_LE}) 
is obtained under the condition that the characteristic
time $\tau_c$ for the force autocorrelation function is negligible on a time scale $\tau_p$
of dynamics of the momentum of the piston.   
Assuming in addition that the random force
in this equation is Gaussian, one can obtain the  
Fokker-Planck equation for the momentum distribution function $f(P)$
\begin{eqnarray}
\frac{\partial f(P)}{\partial t}=nS\,\frac{4m}{M}\,\sqrt{\frac{2k_BT}{\pi m}}
\left\{ \frac{\partial}{\partial P}\Bigl(P f(P)\Bigr) +
M\,k_BT \,\frac{\partial^2f(P)}{\partial P^2}
\right\}.
\end{eqnarray}
This coincides exactly with the equation for 
the piston interacting with the  bath particles
through a hard-wall potential previously obtained by Van Kampen from the
master equation~\cite{VK}.

The assumption of a Gaussian random force appears to be 
justified for Eq. (\ref{normal_LE}) describing 
dynamics on time scales much longer than $\tau_c$. In this case
one can use a coarse-grained description of the dynamics with time resolution
$\tau_c\ll\Delta t\ll\tau_p$. The coarse-graining procedure
corresponds to replacing the instantaneous random force in
Eq.~(\ref{normal_LE}) by its  average over a time window of duration
$\Delta t$, i.e. $F(t)\to \hat{F}(t)=
\Delta t^{-1}\int_t^{t+\Delta t}F(t)\,dt$. As previously discussed,
the number of collisions of bath particles with the piston for the time interval $\tau_p$  
is of order $\lambda^{-2}\gg 1$. Therefore
the resolution time interval $\Delta t$ may be chosen sufficiently long  
that the piston experiences many uncorrelated collisions during $\Delta t$. 
Then, according the central limit theorem, one may expect that $\hat F(t)$
is Gaussian-distributed.

For the more general LE~(\ref{TDLE}) with time-dependent damping, the
random force is generally not Gaussian-distributed.
However,  one can easily demonstrate that the distribution of the
random force is approximately Gaussian in the limit $N \gg 1$ where 
the piston interact simultaneously with many bath particles. 
In fact,  the cumulant expansion~(\ref{cumexp}) of 
the multi-time correlation function 
$C_{2s}=\langle F(t_1)F(t_2)\cdots F(t_{2s})\rangle $
contains the products of $s$ pair correlation functions
$\langle F(t_i)F(t_j)\rangle=\langle\!\langle F(t_i)F(t_j)\rangle\!\rangle$. 
Since a cumulant of any order is proportional $N$, these terms are of order 
$N^s$. The other terms in the expansion contain fewer factors of the
cumulants and therefore fewer factors of $N$,  and hence may be neglected.
Then  $C_{2s}$ can be approximately expressed as a linear combination of pair correlation 
functions, a well-known characteristic of a Gaussian random variable. 
The explicit form of the 
distribution function for the random force $f(F_0)$
can be obtained using  the 
inverse Fourier transformation of the generating
function~(see, for example, reference~\cite{VanKampen}) 
\begin{eqnarray}
f(F_0)=\frac{1}{2\pi}\int_{-\infty}^{\infty}dk\,\exp\left\{
-ikF_0+\sum_{s=1}^{\infty}\frac{(ik)^s}{s!}\langle\!\langle F_0^s\rangle\!\rangle
\right\}.
\label{generation}
\end{eqnarray}
The cumulants of odd orders vanish, and the cumulant of even orders
are $ \langle\!\langle F_0^{2s}\rangle\!\rangle=
2\langle\!\langle F_l^{2s}\rangle\!\rangle$ where $F_l$ is given by 
(\ref{force1}).   Using (\ref{cumdens}) one finds that
\begin{eqnarray}
\langle\!\langle F_l^{2s}\rangle\!\rangle=
\frac{1}{\sqrt{2}}\,N\, g^{2s}\, s!\,\frac{\Gamma(s+1/2)}{\Gamma(s+1)},
\label{cumulants3}
\end{eqnarray}
where $g^2=k_f^2\langle v^2\rangle/2\omega^2$, and $\Gamma(s)$ is the
gamma-function. Substitution (\ref{cumulants3}) in (\ref{generation})
gives the following integral representation for the distribution function
\begin{eqnarray}
f(F)=\frac{1}{2\pi}\int_{-\infty}^{\infty}dk\,\exp\left\{
-ikF-\sqrt{2\pi}\,N\left(1-e^{-(gk)^2}\right)
\right\}
\end{eqnarray}
If $N\gg 1$, one can approximately write in the above expression
$1-e^{-(gk)^2}\approx (gk)^2$, which leads immediately to 
the Gaussian distribution
for the force. The similar arguments hold for the distribution function
of higher order $f\bigl(F(t_1), F(t_2),\cdots F(t_s)\bigr)$.

\section {The nonlinear Langevin equation}
We now turn our attention to the terms of higher order in $\lambda$ in
the $\lambda$-expansion (\ref{expand}) of the memory function
$K(t)\equiv\langle FF^{\dagger}(t)\rangle=\sum_lK_l(t)$. 
It was shown in section 3 that for a homogeneous bath 
$K_1(t)=0$, and  the first non-zero correction to 
$K_0(t)=\langle FF_0(t)\rangle$ is $K_2(t)$ which is of second order in $\lambda$. From 
Eqs.~(\ref{expand}) and (\ref{C2}) one can see that $K_2$ has the structure
\begin{eqnarray}
K_2(t)&=&\frac{\lambda^2}{m^2}A_1(t)P_*^2+\frac{\lambda^2}{m}A_2(t),
\label{K2new}
\end{eqnarray}
where the functions $A_1(t)$ and $A_2(t)$ are given by
\begin{eqnarray}
A_1(t)&=&\int_0^t dt_1\int_0^{t_1} dt_2
\langle\!\langle G_0G_2(t,t_1,t_2)\rangle\!\rangle,\label{A1}\\
A_2(t)&=&\int_0^t dt_1\int_0^{t_1} dt_2
\langle\!\langle G_0G_0(t-t_1)G_1(t,t_2)\rangle\!\rangle\label{A2}.
\end{eqnarray}
Substitution of $K\approx K_0+K_2$ into the exact equation of motion (\ref{EM})
leads to the non-linear generalized LE of the form
\begin{eqnarray}
\frac{dP_*(t)}{dt}=\lambda F^{\dagger}(t) -
{\lambda}^2\int_0^t d\tau\,\,
M_1(\tau) P_*(t-\tau)-
{\lambda}^4\int_0^t d\tau\,\,
M_2(\tau) P_*^3(t-\tau),
\label{NGLEgen}
\end{eqnarray}  
where the memory functions $M_{1}(t)$ and $M_{2}(t)$ are
\begin{eqnarray}
M_1(t)&=&M_0(t)-\frac{2\lambda^2}{m^2} A_1(t)+
\frac{\lambda^2\beta}{m^2}A_2(t),\label{M1}\\
M_2(t)&=&\frac{\beta}{m^3}A_1(t) \label{M2}.
\end{eqnarray}
It is interesting to note that a nonlinear LE of this form has
previously been obtained using a mode
coupling approach~\cite{Ray3}. A similar Markovian version of a
nonlinear Langevin equation with a cubic damping term was considered 
by MacDonald~\cite{MD} on a purely phenomenological grounds. 

Eq.~(\ref{NGLEgen}) differs from the linear 
LE~(\ref{GLE}) with memory function 
$M_0(t)=\frac{\beta}{m}\langle FF_0(t)\rangle$ 
not only by the presence of nonlinear
damping, but also by appearance of correction terms of order $\lambda^2$ 
to the memory function $M_1(t)$ for the linear damping.  Note also that
the last term on the right-hand side of (\ref{NGLEgen}) can be written
in the local form $-\lambda^4P_*^3(t)\int_0^td\tau M_2(\tau)$ since the non-local correction to this expression 
$\lambda^4\int_0^td\tau\,M_2(\tau)\int_{t-\tau}^td\tau'\dot{P_*^3}(\tau')$ 
is of order $\lambda^5$\cite{comment}.  However, the linear damping term in 
Eq.~(\ref{NGLEgen}) cannot be simply written in the local form 
$-\lambda^2P_*(t)\int_0^td\tau M_1(\tau)$ since in this case the non-local 
correction has contributions of  order $\lambda^3$  and $\lambda^4$ which
must be retained.
This correction can be written in the form
\begin{eqnarray}
&&\lambda^2\int_0^t d\tau M_1(\tau)
\int_{t-\tau}^t d\tau'\,\dot{P}_*(\tau')=
\lambda^3\int_0^t d\tau M_0(\tau)
\int_{t-\tau}^t d\tau'\,F^\dagger(\tau')\nonumber\\
&&-\lambda^4\int_0^t d\tau M_0(\tau)
\int_{t-\tau}^t d\tau'\int_0^{\tau'}d\tau''M_0(\tau'')P_*(\tau'-\tau'') 
+O(\lambda^5),
\label{estim2}
\end{eqnarray}
where we have used the result that $M_1(t)=M_0(t)+O(\lambda^2)$ according to (\ref{M1}).
The first term in the right-hand side of this expression  depends on 
initial coordinates of the bath and  may be treated as 
 a small correction to the random force $F^{\dagger}(t)$.
The second term can be written in the local form
\begin{eqnarray}
-\lambda^4P_*(t)\int_0^t d\tau M_0(\tau)
\int_{t-\tau}^t d\tau'\int_0^{\tau'}d\tau''M_0(\tau'')+O(\lambda^5)
\end{eqnarray}
to order $\lambda^5$.
As a result, Eq.~(\ref{NGLEgen}) can be written in the local form
\begin{eqnarray}
\frac{dP_*(t)}{dt}=\lambda \tilde{F}^{\dagger}(t) -
{\lambda}^2\gamma_1(t)P_*(t)-
{\lambda}^4\gamma_2(t)P_*^3(t),
\label{NLEgen}
\end{eqnarray}  
with the modified random force
\begin{eqnarray}
\tilde{F}^{\dagger}(t)=F^{\dagger}(t)+\lambda^3
\int_0^t d\tau M_0(\tau)
\int_{t-\tau}^t d\tau'\,F^\dagger(\tau')
\end{eqnarray}
and the damping functions given by 
\begin{eqnarray}
&&\gamma_1(t)=\int_0^t d\tau M_1(\tau)+\lambda^2
\int_0^t d\tau M_0(\tau)
\int_{t-\tau}^t d\tau'\int_0^{\tau'}d\tau''M_0(\tau''),\label{gamma1}\\
&&\gamma_2(t)=\int_0^t d\tau M_2(\tau).\label{gamma2}
\end{eqnarray}
For $t>>\tau_c$, the time-dependent coefficients $\gamma_i(t)$ attain
their limiting time-independent values $\gamma_i$, which can be obtained
from  the expressions above by setting the upper integration limit $t$ to
infinity.

It is possible to obtained
explicit expressions for the memory functions $M_i(t)$ and
the damping functions $\gamma_i(t)$ for the extended Rayleigh model.  
To accomplish this, explicit expressions for
the functions $G_i$ defined by Eqs.(\ref{Gs}) must be computed.  It is
convenient to express these functions as the sum of two parts corresponding to the 
force acting on left and right  sides of the piston, 
$G_i=G_{li}+G_{ri}$.  It is somewhat problematic to calculate terms
such as $G_{l1}=e^{{\mathcal L}_0\tau}\partial F_l(t)/\partial X$ due to the
parametric dependence of $F_{l}(t)$ on $X$ that is evident when the force is
expressed in terms of $q_i(t)$ and $v_i(t)$ (see Eq.~(\ref{force1})).
One straightforward, albeit inelegant, way to circumvent this difficulty is to
express the force in terms of $N(x,v)\equiv N(x,v;t=0)$. 
Details of this technique can be found in the Appendix B.
Using this approach, we obtain for the left-side part of $G_1$ 
\begin{multline}\label{G1}
G_{l1}(t_1,t_2) =
k_f \, \int\limits_0^\infty dv\int\limits_{-v(\tau_c+t_1)}^{b}dq \,\,
N(X_l+q,v;-\tau_c)\,\,
\cos\omega \left( t_1+\frac{q}{v} \right) \\
\qquad -k_f \theta(\tau_c-t_2)\cos\omega t_2 \int\limits_{0}^{\infty} dv
\int\limits_{-v(\tau_c+t_1-t_2)}^{-vt_1}dq\,\,
N(X_l+q,v;-\tau_c), 
\end{multline}
where the integration limit $b$  is
\begin{eqnarray}
b=-v(\tau_c+t_1-t_2)\theta(\tau_c-t_2)-vt_1\theta(t_2-\tau_c).
\end{eqnarray}
The expression for $G_{r1}$ can be obtained
from the one above by replacements,
\begin{equation}
\int_{0}^{\infty}dv\to \int_{-\infty}^{0} dv,\,\,\,\,\,
\int_{q_1}^{q_2}dq\to \int_{q_2}^{q_1} dq.
\label{replace}
\end{equation}
Note that 
$\langle G_{l1}\rangle=\langle G_{r1}\rangle$.

For the left-side contribution to $G_2$, we obtain
\begin{multline}\label{G2}
G_{l2}(t,t_1,t_2)=
k_f \, \int\limits_0^\infty dv \,\,N(X_l-v(t+\tau_c),v;-\tau_c)\\
+k_f \, \int\limits_{0}^{\infty} dv\int\limits_{-v(t+\tau_c)}^{b}dq\,\,
N(X_l+q,v;-\tau_c)\frac{\omega}{v}
\sin\omega\left( t+\frac{q}{v} \right)\\
+k_f \phi(t_1,t_2)\int\limits_0^\infty dv \,\,N(X_l-vt,v;-\tau_c),
\end{multline}
where the function $\phi(t,t_1,t_2)$ is
\begin{eqnarray}
\phi(t_1,t_2) &=& \theta(t_2-\tau_c)-
\theta(t_1-\tau_c)\theta(\tau_c-t_2)\cos\omega t_2\nonumber\\
&+&\theta(\tau_c-t_1)\theta(\tau_c-t_2)\cos\omega t_2\cos\omega t_1,
\end{eqnarray}
and the upper integration limit $b$ is 
\begin{eqnarray}
b=-v(\tau_c+t-t_2)\theta(\tau_c-t_2)-vt\theta(t_2-\tau_c).
\end{eqnarray} 

A similar expression for $G_{r2}$ can be obtained from that for $G_{l2}$
using replacements (\ref{replace}), and, in addition, by
multiplying the first and the last terms on the right-hand side of
Eq.~(\ref{G2}) by $-1$.
Recall that we anticipate $\langle G_{2}\rangle=0$ by symmetry so that
$\langle G_{r2}\rangle=-\langle G_{l2}\rangle$, which can be 
explicitly verified from the expressions above.

It has been shown above that
the damping forces in the nonlinear Langevin equation 
can be expressed as integrals of the
cumulants $\langle\!\langle G_{0}G_{2}\rangle\!\rangle$
and $\langle\!\langle G_{0}G_{0}G_1\rangle\!\rangle$.
From Eqs.~(\ref{G1}), (\ref{G2}), (\ref{force1}), and ~(\ref{cumdens}),
one can get an explicit expressions for
these cumulants. For $t>t_1>t_2$ we find
\begin{eqnarray}
\langle\!\langle G_{0}G_{2}(t,t_1,t_2)\rangle\!\rangle &=&
2\langle\!\langle G_{l0}G_{l2}(t,t_1,t_2)\rangle\!\rangle\nonumber\\
&=&\frac{nSk_{f}^2}{\omega}\phi_1(t,t_1,t_2)\int_0^{\infty}dv\,f_M(v)v,
\end{eqnarray}
where 
\begin{eqnarray}
\phi_1(t,t_1,t_2)=2\theta(\tau_c-t)\sin\omega t\cos\omega t_1\cos\omega 
t_2.
\end{eqnarray}
The second cumulant required  is
\begin{eqnarray}
\langle\!\langle G_{0}G_{0}(t-t_1)G_1(t,t_2)\rangle\!\rangle&=& 
2\langle\!\langle G_{l0}G_{l0}(t-t_1)G_{l1}(t,t_2)\rangle\!\rangle
\nonumber\\
&=&-nS\left(\frac{k_{f}}{\omega}\right)^3\phi_2(t,t_1,t_2)
\int_0^{\infty}dv\,f_M(v)v^3 \nonumber,
\end{eqnarray} 
where the function $\phi_2$ has the form
\begin{multline*}
\phi_2(t,t_1,t_2)=\theta(\tau_c-t)\cos\omega t_2\times \\
\Bigl\{\omega(\tau_c-t)\cos\omega(t-t_1)+
\frac{1}{2}\sin\omega(t+t_1)+\frac{1}{2}\sin\omega(t-t_1)\Bigr\}.
\end{multline*} 

These results allow to calculate the functions $A_i(t)$ defined by
Eqs.~(\ref{A1}) and (\ref{A2}),
\begin{eqnarray}
A_1(t)&=&m^2\omega^2N\,\xi_1(t),\\
A_2(t)&=&-m^2\omega^2\beta^{-1}N\,\xi_2(t),
\end{eqnarray}
where dimensionless functions $\xi_i(t)$ are given by
\begin{eqnarray}
\xi_1(t)&=&\frac{1}{\sqrt{2\pi}}\theta(\tau_c-t)\sin^3\omega t,\nonumber\\
\xi_2(t)&=&\frac{1}{\sqrt{2\pi}}\theta(\tau_c-t)
\Bigl\{\sin^3\omega t+\omega t(\pi-\omega t)\sin\omega t\Bigr\}.
\end{eqnarray}
The memory functions in the non-Markovian 
LE equation~(\ref{NGLEgen}), take the form
\begin{eqnarray}
M_1(t)&=&N\omega^2\Bigl\{\xi_0(t)-2\lambda^2\xi_1(t)-
\lambda^2\xi_2(t)\Bigr\},\\
M_2(t)&=&N\omega^2\frac{\beta}{m}\xi_1(t),
\end{eqnarray}
Here the function $\xi_0(t)$ is given by (\ref{xi0})  and governs
the memory function $M_0(t)$ in the linear LE (\ref{GLE}), 
$M_0(t)=N\omega^2\xi_0(t)$. 
The additional $\lambda^2$-correction terms leads to a faster decay of 
the linear damping kernel $M_1(t)$ compared to $M_0(t)$.
The kernel for nonlinear damping $M_2(t)$  
is not decaying function of time but rather has a maximum at $t=\tau_c/2$.

Explicit expressions for damping functions $\gamma_i(t)$ 
in the local-in-time  LE (\ref{NLEgen}) by integrating $M_i(t)$
according to Eqs. (\ref{gamma1}) and (\ref{gamma2}), and the nonlinear LE 
for the Rayleigh model with a parabolic 
potential takes  the form
\begin{eqnarray}
\frac{dP_*(t)}{dt}=\lambda \tilde{F}^\dagger(t)-
N{\lambda}^2\omega \Bigl\{\zeta_1(t)P_*(t)+
\lambda^2\frac{\beta}{m}\zeta_2(t)P_*^3(t)\Bigr\}, 
\label{NLE}
\end{eqnarray}
where the nonlinear damping function $\zeta_2(t)$ is given by 
\begin{eqnarray}
\zeta_2(t)=\omega\int_0^t d\tau\xi_1(\tau)=\frac{1}{\sqrt{2\pi}}
\theta(\tau_c-t)\left\{\frac{2}{3}-\cos\omega t+\frac{1}{3}\cos^3\omega t
\right\}
+\frac{1}{3}\sqrt{\frac{8}{\pi}}\theta(t-\tau_c),
\end{eqnarray}
and the linear damping function $\gamma_1(t)$  can be written as
\begin{eqnarray}
\zeta_1(t)&=&\zeta_0(t)+\lambda^2\varepsilon_1(t)+
N\lambda^2\,\varepsilon_2(t).
\end{eqnarray}
The main contribution to the overall damping coefficient $\zeta_1(t)$
is given by the function  
$\zeta_0(t)=\omega\int_0^t d\tau\xi_0(\tau)$, while the corrections
arising at higher orders in the $\lambda$-expansion are
\begin{eqnarray}
\varepsilon_1(t)&=&-\omega\int_0^t d\tau\Bigl\{2\xi_1(\tau)+
\xi_2(\tau)\Bigr\},\\
\varepsilon_2(t)&=&\omega^3\int_0^t d\tau\xi_0(\tau)
\int_{t-\tau}^{t}d\tau'\int_0^{\tau'}d\tau''\xi_0(\tau'').
\end{eqnarray}
Note that $\varepsilon_1(t)$ and $\varepsilon_2(t)$ are of different signs
for all $t$. 
If $N\ll 1$ the correction is determined mostly by $\varepsilon_1(t)<0$ 
and tends to decrease the damping function $\zeta_1(t)$. 
In contrast, if
$N\gg 1$ the main correction comes from $\varepsilon_2(t)>0$
effectively increasing the linear damping. 

For a coarse-grained description on the time-scale $t\gg \tau_c$ 
with a time resolution
$\tau_c\ll\Delta t\ll \tau_p$, one can replace the damping 
functions $\gamma_i(t)$ in the LE (\ref{NLE}) by their limiting values 
$\gamma_i(\tau_c)$.

\section{Concluding remarks}
The notion that the character of Brownian motion of a finite-sized particle
may depend on parameters other than the mass ratio $\lambda^2$ dates
back to Lorentz and has been examined by many authors 
(see~\cite{Tokuyama,Roux,Bocquet} and references therein).    
It is known that when hydrodynamic effects are important
another relevant parameter is the ratio of the mass
density of the bath to that of the particle. 
In this paper we have
demonstrated that even when hydrodynamic effects are absent, as in the extended
Rayleigh model, the character of the behavior of a tagged particle   
may  not governed by $\lambda^2$ but by the renormalized parameter
$\lambda_*^2=N\lambda^2$, which can be interpreted as the ratio of
the average total mass of particles in the interaction zone $M_*$ 
to the mass $M$ of the tagged particle. 
When the average number of particles in the interaction zone
is large (i.e. $N\gg 1$),  $\lambda_* \ll 1$ is a 
necessary condition for the applicability of a conventional perturbation 
scheme of derivation of the LE. In this case the conventional assumption
of Gaussian random force is justified for any time scale. 
If $N<1$, the parameter of the expansion is $\lambda^2$, and the
Gaussian force approximation holds only on a time scale that
is much longer than the characteristic time for the relaxation of the bath. 

Although this paper focuses on the specific model of an ideal gas bath 
interacting with a Brownian particle through a quadratic repulsive potential,
many of the results obtained are quite general.  In particular, the LE (\ref{NLEgen}) and  the expressions 
(\ref{gamma1})-(\ref{gamma2}) for the damping coefficients in terms
of microscopic time-correlation functions, which may be considered as the 
generalized version of the fluctuation-dissipation theorem, are
limited neither to the specific form of interaction potential between
the bath and the tagged particle nor to the ideal gas bath. The results of
section 3 concerning the cumulant expansion of the kernel 
$K(t)=\langle FF^{\dagger}(t)\rangle$ are also general and not limited 
to any specific model. Combined with quite general theorems about 
cumulant properties~\cite{SO}, these results may be useful 
for more realistic models with  interacting bath particles.

The explicit expression for the kinetic coefficients and
memory functions appearing in the Langevin equations have been derived 
in this paper in the thermodynamic limit, so that any correlations due 
to finite size of the system are neglected.
It should be mentioned, however, that the equations themselves, 
as well as the fluctuation-dissipation relations relating the kinetic coefficients 
to correlation functions, also hold for a system with finite baths.
For finite systems the explicit form of the kinetic
coefficients may be rather complicated even for a bath
composed of ideal gas particles.

In this paper we have considered the case of the totally symmetric bath
when thermodynamic and microscopic
properties of the gas to the left and to the right of the piston 
are the same.
Some interesting physical 
implications arising in the case of an asymmetric bath 
will be presented elsewhere~\cite{PS}.

\section*{Acknowledgments}
This work was supported by a grant from the
Natural Sciences and Engineering Research Council of Canada and funds
from the Premier's Research Excellence Award.

\renewcommand{\theequation}{A\arabic{equation}}
\setcounter{equation}{0} 
\section*{Appendix A}
In this Appendix, for the sake of completeness,
 we show explicitly that the function $C_4$ 
defined by Eqs.~(\ref{CF}) is of second order in cumulants of products
of $G_i$. $C_{4}$ can be written as the sum
\begin{eqnarray} 
C_4=C_4^{(1)}+C_4^{(2)}+C_4^{(3)}+C_4^{(4)},
\nonumber
\end{eqnarray}
where the first constituent,
\begin{multline*}
C_4^{(1)}=\frac{2}{m^2}\Bigl\{\langle G_0G_0(t-t_1)G_0(t-t_2)G_2(t,t_3,t_4)
\rangle \\
-\langle G_0G_0(t-t_1)\rangle\langle G_0(t-t_2)G_2(t,t_3,t_4)\rangle
\Bigr\} ,
\end{multline*}
is obviously quadratic in cumulants,
\begin{eqnarray*}
C_4^{(1)} &=& \frac{2}{m^2}\Bigl\{\langle\!\langle G_0G_0(t-t_2)\rangle\!\rangle
\langle\!\langle G_0(t-t_1)G_2(t,t_3,t_4)\rangle\!\rangle \\
&&+ \langle\!\langle G_0(t-t_1)G_0(t-t_2)\rangle\!\rangle
\langle\!\langle G_0G_2(t,t_3,t_4)\rangle\!\rangle \\
&&+ \langle\!\langle 
G_0G_0(t-t_1)G_0(t-t_2)G_2(t,t_3,t_4)\rangle\!\rangle\Big\}.
\end{eqnarray*}
The second term is
\begin{eqnarray*}
C_4^{(2)} &=&
\frac{1}{m^2}\Bigl\{\langle G_0G_0(t-t_1)A_1 \rangle-
\langle G_0G_0(t-t_1)\rangle\langle A_1\rangle \\
&&- \langle G_1(t,t_4)\rangle\Bigl[\langle G_0G_0(t-t_1)
G_1(t-t_3,t_2-t_3)\rangle \\
&&-\langle G_0G_0(t-t_1)\rangle\langle G_1(t-t_3,t_2-t_3)
\rangle\Bigr]\Bigr\},
\end{eqnarray*}
where $A_1=S(t-t_2)G_0(t_2-t_3)G_1(t_2,t_4)$. 
Noting that $A_1$ can be written as
\begin{eqnarray}
A_1=G_1(t-t_3,t_2-t_3)G_1(t,t_4)+G_0(t-t_3)G_2(t,t_2,t_4),
\nonumber
\end{eqnarray}
and recalling that due to symmetry $\langle G_2(t)\rangle$ and 
$\langle G_0G_1(t)\rangle$ are zero at all times  (see Eq.~(\ref{sym})),
one can see that $C_4^{(2)}$ is also quadratic in cumulants,
\begin{eqnarray*}
C_4^{(2)} &=&
\frac{1}{m^2}\Bigl\{
\langle G_1(t-t_3,t_2-t_3)\rangle
\langle\!\langle G_0G_0(t-t_1)G_1(t,t_4)\rangle\!\rangle \\
&&+\langle\!\langle G_0G_0(t-t_3)\rangle\!\rangle
\langle\!\langle G_0(t-t_1)G_2(t,t_2,t_4)\rangle\!\rangle \\
&&+\langle\!\langle G_0G_2(t,t_2,t_4)\rangle\!\rangle
\langle\!\langle G_0(t-t_1)G_0(t-t_3)\rangle\!\rangle \\
&&+\langle\!\langle G_0G_0(t-t_1)G_1(t-t_3,t_2-t_3)
G_1(t,t_4)\rangle\!\rangle \\
&&+\langle\!\langle G_0G_0(t-t_1)G_0(t-t_3)G_2(t,t_2,t_4)
G_1(t,t_4)\rangle\!\rangle\Bigr\}.
\end{eqnarray*}
The third term is
\begin{eqnarray}
C_4^{(3)}=\frac{P_*^2}{m^3}\Bigl\{\langle G_0 B_1\rangle-
\langle G_0B_2\rangle
\langle G_1(t_2,t_4)\rangle+2\langle G_0B_3\rangle\Bigr\},\nonumber 
\end{eqnarray}
where
\begin{eqnarray*}
B_1&=&S(t-t_1)S(t-t_2)G_0(t_2-t_3)G_1(t_2,t_4), \\
B_2&=&S(t-t_1)S(t-t_2)G_0(t_2-t_3),\\
B_3&=&S(t-t_1)G_0(t_1-t_2)G_2(t_1,t_3,t_4),
\end{eqnarray*}
can be expressed as
\begin{align*}
B_1 &=G_2(t-t_3,t_1-t_3,t_2-t_3)G_1(t,t_4)+
G_1(t-t_3,t_2-t_3)G_2(t,t_1,t_4)\\
& +G_1(t-t_3,t_1-t_3)G_2(t,t_2,t_4)+
G_0(t-t_3)G_3(t,t_1,t_2,t_4),\\
B_2 &=G_2(t-t_3,t_1-t_3,t_2-t_3), \\
B_3 &=G_1(t-t_2,t_1-t_2)G_2(t,t_3,t_4)+
G_0(t-t_2)G_3(t,t_1,t_3,t_4) .
\end{align*}
Then using the symmetry property (\ref{sym}) one can see that 
$C_4^{(3)}$ is quadratic in cumulants.

The remaining term
\begin{eqnarray*}
C_4^{(4)} &=&\left(\frac{P_*}{m}\right)^2
\langle G_0G_4(t,t_1,t_2,t_3,t_4)\rangle \\
&&+\frac{3P_*^2}{m^3}\Bigl\{\langle G_0G_0(t-t_1)
G_3(t,t_2,t_3,t_4)\rangle \\
&&\quad -\langle G_0G_0(t-t_1)\rangle\langle G_3(t,t_2,t_3,t_4)\rangle
\Bigr\}
\end{eqnarray*}
is clearly linear in cumulants,
\begin{eqnarray*}
C_4^{(4)}&=&\left(\frac{P_*}{m}\right)^2
\langle\!\langle G_0G_4(t,t_1,t_2,t_3,t_4)\rangle\!\rangle \\
&&+ \frac{3P_*^2}{m^3}\langle\!\langle G_0G_0(t-t_1)
G_3(t,t_2,t_3,t_4)\rangle\!\rangle.
\end{eqnarray*}

\renewcommand{\theequation}{B\arabic{equation}}
\setcounter{equation}{0}
\section*{Appendix B}
In this Appendix the functions $G_i$ 
defined by Eqs.~(\ref{Gs}) are evaluated 
for the extended Rayleigh model of diffusion. These functions
are defined in terms of powers of alternating operators $\partial/\partial X$
and $e^{{\mathcal L}_0(t_i-t_k)}$. As mentioned in the text, the 
representation (\ref{force1}), (\ref{force11}) for the force on the fixed piston 
is unwieldy since it involves  $q_i(t)$
and $v_i(t)$ which are functions of $X$.  For the purpose of
evaluating the $G_i$, it is convenient to
express the force in terms of $N(x,v)\equiv N(x,v;t=0)$. For $t>0$,
we have
\begin{eqnarray}
F_l(t) &=& k_f\, \int\limits_0^\infty dv\int\limits_{-vt}^{a(t)}dq \,\,
N(X_l+q,v)\,\,\,\frac{v}{\omega}
\sin\omega \left( t+\frac{q}{v} \right) \nonumber \\
&&+\theta(\tau_c/2-t)
k_f\int\limits_0^\infty dv\int\limits_{0}^{\infty}dq \,\,
N(X_l+q,v)\,\,q(t) \nonumber\\
&&+\theta(\tau_c/2-t)
k_f\int\limits_{-\infty}^{0} dv\int\limits_{Q(t)}^{\infty}dq \,\,
N(X_l+q,v)\,\,q(t)\nonumber\\
&&+\theta(t-\tau_c/2)\theta(\tau_c-t)
k_f\int\limits_0^\infty dv\int\limits_{0}^{Q(t)}dq \,\,
N(X_l+q,v)\,\,q(t),
\label{A21}
\end{eqnarray} 
where
\begin{eqnarray}
a(t)&=&-v\theta(t-\tau_c)(t-\tau_c),\nonumber\\
q(t)&=&q\cos\omega t+\frac{v}{\omega}\sin\omega t,\nonumber\\ 
Q(t)&=&-\frac{v}{\omega}\tan\omega t.\nonumber
\end{eqnarray}

The first term in Eq.~(\ref{A21})
describes the contribution to the force $F_l(t)$ from the particles that
were outside the interaction zone at $t=0$ (i.e. $q<0$) and are in the interaction zone at the moment $t$
(i.e. $q(t)>0$). 
 
The remaining terms give the contribution from
particles that were in the interaction zone at $t=0$ (i.e. $q>0$)
and are still there at the moment $t$ (i.e. $q(t)>0$). In fact,
all particles in the interaction zone with positive initial velocities at $t=0$ 
will be still in the interaction zone at $t<\tau_c/2$ (the second term), while the particles
with negative initial velocities will be in the interaction zone at time $t<\tau_c/2$ only 
if at $t=0$  they reside deep inside the interaction zone, namely $q>Q(t)$ (the third term). 
For $\tau_c>t>\tau_c/2$ only the particles with positive initial velocity will
be in the interaction zone at time $t$ provided their initial coordinates  are less than
$Q(t)$ (the last term).

From expression (\ref{A21}), one easily calculates $\partial F_l(t)/\partial X$
writing $\partial N(X_l+q,v)/\partial X=\partial N(X_l+q)/\partial q$
and integrating by parts to obtain,
\begin{eqnarray}
\frac{\partial F_l(t)}{\partial X} &=& -k_f\cos\omega t N_z(t) \label{A22}\\
&&-k_f\int\limits_0^\infty dv\int\limits_{-vt}^{a(t)}dq \,\,
N(X_l+q,v)\,\,\,\frac{v}{\omega}
\cos\omega \left( t+\frac{q}{v} \right) .\nonumber
\end{eqnarray}
Here $N_z(t)$ is the number of particles which were in the interaction zone at $t=0$
and remain at $t>0$,
\begin{eqnarray}
N_z(t) &=&
\theta(\tau_c/2-t)\int\limits_0^\infty dv\int\limits_{0}^{\infty}dq \,\,
N(X_l+q,v) \nonumber\\
&& + \theta(\tau_c/2-t)
\int\limits_{-\infty}^{0} dv\int\limits_{Q(t)}^{\infty}dq \,\,
N(X_l+q,v) \nonumber\\
&&+\theta(t-\tau_c/2)\theta(\tau_c-t)
\int\limits_0^\infty dv\int\limits_{0}^{Q(t)}dq \,\,
N(X_l+q,v).
\label{Nz1}
\end{eqnarray}
$N_z(t)$ can be written in compact form in terms of the density at time
$-\tau_c$ according to,
\begin{equation}
N_z(t)=\theta(\tau_c-t)\int\limits_{0}^{\infty} dv\int\limits_{-v\tau_c}^{-vt}dq\,\,
N(X_l+q,v;-\tau_c).
\label{Nz2}
\end{equation}
In fact, the number of particles in the interaction zone at $t=0$ is
\begin{equation}
\int\limits_{0}^{\infty} dv\int\limits_{-v\tau_c}^{0}dq\,\,
N(X_l+q,v;-\tau_c),
\label{A24}
\end{equation}
while at time $t$ it is given by
\begin{multline}
\int\limits_{0}^{\infty} dv\int\limits_{-v\tau_c}^{0}dq\,\,
N(X_l+q,v;t-\tau_c) \\
=\int\limits_{0}^{\infty} dv\int\limits_{-v(t+\tau_c)}^{-vt}dq\,\,
N(X_l+q,v;-\tau_c).
\label{A25}
\end{multline}
By definition $N_z(t)$ involves the particles which contribute
to both integrals (\ref{A24}) and (\ref{A25}), which leads to Eq.~(\ref{Nz2}).
Combining Eqs.~(\ref{A22}) and (\ref{Nz2}), we have
\begin{eqnarray}
\frac{\partial F_l(t)}{\partial X} &=&
-k_f\int\limits_0^\infty dv\int\limits_{-vt}^{a(t)}dq \,\,
N(X_l+q,v)\,\,\,
\cos\omega \left( t+\frac{q}{v} \right)\nonumber\\
&&-k_f\theta(\tau_c-t)\cos\omega t
\int\limits_{0}^{\infty} dv\int\limits_{-v\tau_c}^{-vt}dq\,\,
N(X_l+q,v;-\tau_c).
\label{A26} 
\end{eqnarray}
 
For comparison, let us calculate
the average derivative of the total force $F_0(t)=F_l+F_r$, 
\begin{multline}
\left\langle\frac{\partial F_0(t)}{\partial X}\right\rangle =
2\left\langle\frac{\partial F_l(t)}{\partial X}\right\rangle \nonumber
\\ = -\frac{2k_fnS}{\omega}\int_0^\infty f_M(v)v\,dv \theta(\tau_c-t)\Bigl( \sin\omega t+
(\pi-\omega t)\cos\omega t\Bigr).
\end{multline}
Comparing this expression with Eq.~(\ref{K00}) for the memory kernel
$K_0(t)=\langle FF_0(t)\rangle$,  it is evident that
$\langle\partial F_0(t)/\partial X\rangle=-\beta\langle\ FF_0(t)\rangle$.
This is the general result used in the main text, Eq.~(\ref{gradient}) confirmed
using the explicit expression for $\partial F_0(t)/\partial X$.

>From Eq.~(\ref{A26}), we obtain the following expression for 
$G_{l1}$:
\begin{multline}\label{A28}
G_{l1}(t_1,t_2)\equiv e^{{\mathcal L}_0(t_1-t_2)}\frac{\partial
F_l(t_2)}{\partial X} \\
=-k_f\int\limits_0^\infty dv\int\limits_{-vt_2}^{a(t_2)}dq \,\,
N(X_l+q,v;t_1-t_2)\,\,\,
\cos\omega \left( t_2+\frac{q}{v} \right) \\
-k_f\theta(\tau_c-t_2)\cos\omega t_2 \int\limits_{0}^{\infty} dv\int\limits_{-v\tau_c}^{-vt_2}dq\,\,
N(X_l+q,v;t_1-t_2-\tau_c),
\end{multline}
where, according to (\ref{expand}), it is assumed  that $t_1>t_2$.
Expressing this in terms of the microscopic density at time $-\tau_c$
using (\ref{property}), 
one obtains Eq.~(\ref{G1}) of the main text.

To evaluate $G_2$, one has take the derivative  of $G_1$ with respect to $X$.  
Let us express $G_1$ in terms of the density at time $t=0$, $N(x,v)$, 
as done above for $F_l(t)$.
The first term in the right-hand side of Eq.~(\ref{A28}) involves 
only particles located outside the interaction zone, so using the 
property~(\ref{property}), the first term can be written as 
\begin{multline}\label{A29}
-k_f\int\limits_0^\infty dv\int\limits_{-vt_2}^{a(t_2)}dq \,\,
N(X_l+q-v(t_1-t_2),v)\,\,
\cos\omega \left( t_2+\frac{q}{v} \right) \\
=-k_f\int\limits_0^\infty dv\int\limits_{-vt_1}^{b}dq \,\,
N(X_l+q,v)\,\,
\cos\omega \left( t_1+\frac{q}{v} \right),
\end{multline}
where the upper integration limit is
\begin{eqnarray}
b=-v(t_1-t_2)\theta(\tau_c-t_2)-v(t_1-\tau_c)\theta(t_2-\tau_c)\nonumber.
\end{eqnarray}
The second term on the right-hand side of Eq.~(\ref{A28}) can be
expressed as
\begin{multline}
-k_f\theta(\tau_c-t_2)\cos\omega t_2 \int\limits_{0}^{\infty} dv
\int\limits_{-v\tau_c}^{-vt_2}dq\,\,N(X_l+q-v(t_1-t_2),v;-\tau_c) \\
=-k_f\theta(\tau_c-t_2)\cos\omega t_2 \int\limits_{0}^{\infty} dv
\int\limits_{-v(\tau_c+t_1-t_2)}^{-vt_1}dq\,\,N(X_l+q,v;-\tau_c)
\label{A29aux}.
\end{multline}
If $t_1>\tau_c$, then all particles which contribute this integral 
at $t=0$ will be outside the interaction zone in the $q$-interval from
$-v(t_1-t_2)$ to $-v(t_1-\tau_c)$. Therefore, for $t_1>\tau_c$ the second term
 equals
\begin{eqnarray}
-k_f\theta(\tau_c-t_2)\theta(t_1-\tau_c)\cos\omega t_2 \int\limits_{0}^{\infty} dv
\int\limits_{-v(t_1-t_2)}^{-v(t_1-\tau_c)}dq\,\,N(X_l+q,v).
\end{eqnarray}  
If $t_1<\tau_c$, then
two sets of particles contribute to expression (\ref{A29aux}). 
The first set of particles is composed of 
particles which at time $-\tau_c$ are in the $q-$interval 
from $-v(\tau_c+t_1-t_2)$ to $-v\tau_c$. At $t=0$, these particles will be 
outside the interaction zone in the interval $(-v(t_1-t_2),0)$, and hence their contribution
is 
\begin{eqnarray}
-k_f\cos\omega t_2 \int\limits_{0}^{\infty} dv
\int\limits_{-v(t_1-t_2)}^{0}dq\,\,N(X_l+q,v).
\label{A210}
\end{eqnarray}
The second group of particles are those that were in the $q-$interval
from $-v\tau_c$ to $-vt_1$ at time $-\tau_c$. At $t=0$, all these particles will be in the interaction zone.
Taking into account Eq.~(\ref{Nz2}), 
the corresponding contribution can be written as 
$-k_f\cos\omega t_2N_z(t_1)$. Using expression~(\ref{Nz1}) for $N_z(t)$,
one arrives at the following representation for $G_{l1}$ in terms of $N(x,v)$,  
\begin{eqnarray}
G_{l1}(t_1,t_2) &=&
-k_f\int\limits_0^\infty dv\int\limits_{-vt_1}^{b}dq \,\,
N(X_l+q,v)\,\,
\cos\omega \left( t_1+\frac{q}{v} \right) \nonumber\\
&&-k_f\theta(\tau_c-t_2)\theta(t_1-\tau_c)\cos\omega t_2 \int\limits_{0}^{\infty} dv
\int\limits_{-v(t_1-t_2)}^{-v(t_1-\tau_c)}dq\,\,N(X_l+q,v) \nonumber\\
&&-k_f\theta(\tau_c-t_2)\theta(\tau_c-t_1)\cos\omega t_2\int\limits_{0}^{\infty} dv
\int\limits_{-v(t_1-t_2)}^{0}dq\,\,N(X_l+q,v) \nonumber\\
&&-k_f\theta(\tau_c-t_2)\theta(\tau_c/2-t_1)\cos\omega t_2\int\limits_{0}^{\infty} dv
\int\limits_{0}^{\infty}dq\,\,N(X_l+q,v) \nonumber\\
&&-k_f\theta(\tau_c-t_2)\theta(\tau_c/2-t_1)\cos\omega t_2\int\limits_{-\infty}^{0} dv
\int\limits_{Q(t_1)}^{\infty}dq\,\,N(X_l+q,v) \nonumber\\
&&-k_f\theta(\tau_c-t_2)\theta(\tau_c-t_1)\theta(t_1-\tau_c/2)
\cos\omega t_2\times\nonumber\\
&&\quad \quad \int\limits_{0}^{\infty} dv
\int\limits_{0}^{Q(t_1)}dq\,\,N(X_l+q,v) .
\end{eqnarray}
Taking derivative of this expression with respect to $X$ gives
\begin{eqnarray}
\frac{\partial}{\partial X}G_{l1}(t_1,t_2) &=&
k_f\int\limits_0^\infty dv\,\,N(X_l-vt_1,v) \\
&&+k_f[\theta(t_2-\tau_c)-\theta(\tau_c-t_2)\theta(t_1-\tau_c)\cos\omega t_2]
\int\limits_0^\infty dv\,\,N(X_l-v(t_1-\tau_c),v) \nonumber\\
&&-k_f\int\limits_0^\infty dv\int\limits_{-vt_1}^{b}dq \,\,
N(X_l+q,v)\,\,\frac{\omega}{v}
\sin\omega \left( t_1+\frac{q}{v} \right) \nonumber\\
&&+ \theta(\tau_c/2-t_1)\theta(\tau_c-t_2)k\cos\omega t_2 \int\limits_{-\infty}^0 dv
\,\,N(X_l+Q(t_1),v) \nonumber\\
&&-\theta(\tau_c-t_1)\theta(t_1-\tau_c/2)\theta(\tau_c-t_2)k\cos\omega t_2 \int\limits_0^{\infty} dv
\,\,N(X_l+Q(t_1),v). \nonumber
\label{A212}
\end{eqnarray}
The last two terms can be written as
\begin{multline}\label{A213}
k_f\theta(\tau_c-t_2)\theta(\tau_c-t_1)\frac{\cos\omega t_2}{\tan\omega t_1}
\int\limits_0^{\infty} dv
\,\,N\left(X_l+\frac{v}{\omega},-\frac{v}{\tan\omega t_1}\right) \\
=k_f\theta(\tau_c-t_2)\theta(\tau_c-t_1)\frac{\cos\omega t_2}{\tan\omega t_1}
\int\limits_0^{\infty} dv
\,\,N\left(X_l-\frac{vt_1}{\sin\omega t_1},\,\frac{v}{\sin\omega
t_1};-\tau_c\right) \\
=k_f\theta(\tau_c-t_2)\theta(\tau_c-t_1)\cos\omega t_1\cos\omega t_2\int\limits_0^{\infty} dv
\,\,N\left(X_l-vt_1,v;-\tau_c\right),
\end{multline}
where we have used the property that if the initial coordinate and velocity of the 
particle in the interaction zone are
\begin{eqnarray}
q(0)=\frac{V}{\omega},\,\,\,v(0)=-\frac{V}{\tan\omega t} 
\end{eqnarray}
with $V>0$ and $0<t<\tau_c$, then at $t=-\tau_c$
\begin{eqnarray}
q(-\tau_c)=-v(-\tau_c)t,\qquad v(-\tau_c)=\frac{V}{\sin\omega t}.
\end{eqnarray}

Substituting (\ref{A213}) into (\ref{A212}), acting on the result by 
the propagator $e^{{\mathcal L}_0(t-t_1)}$, and using again the property
$N(x,v;t+\tau)=N(x-v\tau,v;t)$ for the motion outside the interaction zone, we finally
obtain for $t>t_1>t_2$,
\begin{eqnarray}
G_{l2}(t,t_1,t_2) &\equiv&  
e^{{\mathcal L}_0(t-t_1)}\frac{\partial}{\partial X}G_1(t_1,t_2) \nonumber\\
&=& k_f\int\limits_0^\infty dv \,\,N(X_l-vt,v) \\
&&+k_f\theta(t_1-\tau_c)\theta(t_2-\tau_c)\int\limits_0^\infty dv
\,\,N(X_l-v(t-\tau_c),v) \nonumber \\
&&-k_f\theta(\tau_c-t_2)\theta(t_1-\tau_c)\cos\omega t_2
\int\limits_0^\infty dv \,\,N(X_l-v(t-\tau_c),v) \nonumber\\
&&-k_f\int\limits_{0}^{\infty} dv\int\limits_{-vt}^{b}dq\,\,
N(X_l+q,v)\frac{\omega}{v}
\sin\omega\left( t+\frac{q}{v} \right) \nonumber\\
&&+k_f\theta(\tau_c-t_2)\theta(\tau_c-t_1)\cos\omega t_1\cos\omega t_2
\int\limits_0^\infty dv \,\,N(X_l-vt,v;-\tau_c), \nonumber
\label{A2G2}
\end{eqnarray}
where $b=-v(t-t_2)\theta(\tau_c-t_2)-v(t-\tau_c)\theta(t_2-\tau_c)$.
Expressing the first three terms  through 
$N(x,v;-\tau_c)$ rather than $N(x,v)$, one arrives at Eq.~(\ref{G2})
of the main text.



\begin{thebibliography}{9}
\bibitem{identity} A simple way to prove this formula is to
integrate the identity 
$\frac{d}{dt}e^{{\mathcal A}t}e^{-({\mathcal A+B})t}=-e^{{\mathcal
A}t}{\mathcal B}e^{-({\mathcal A+B})t}$.
\end{thebibliography}
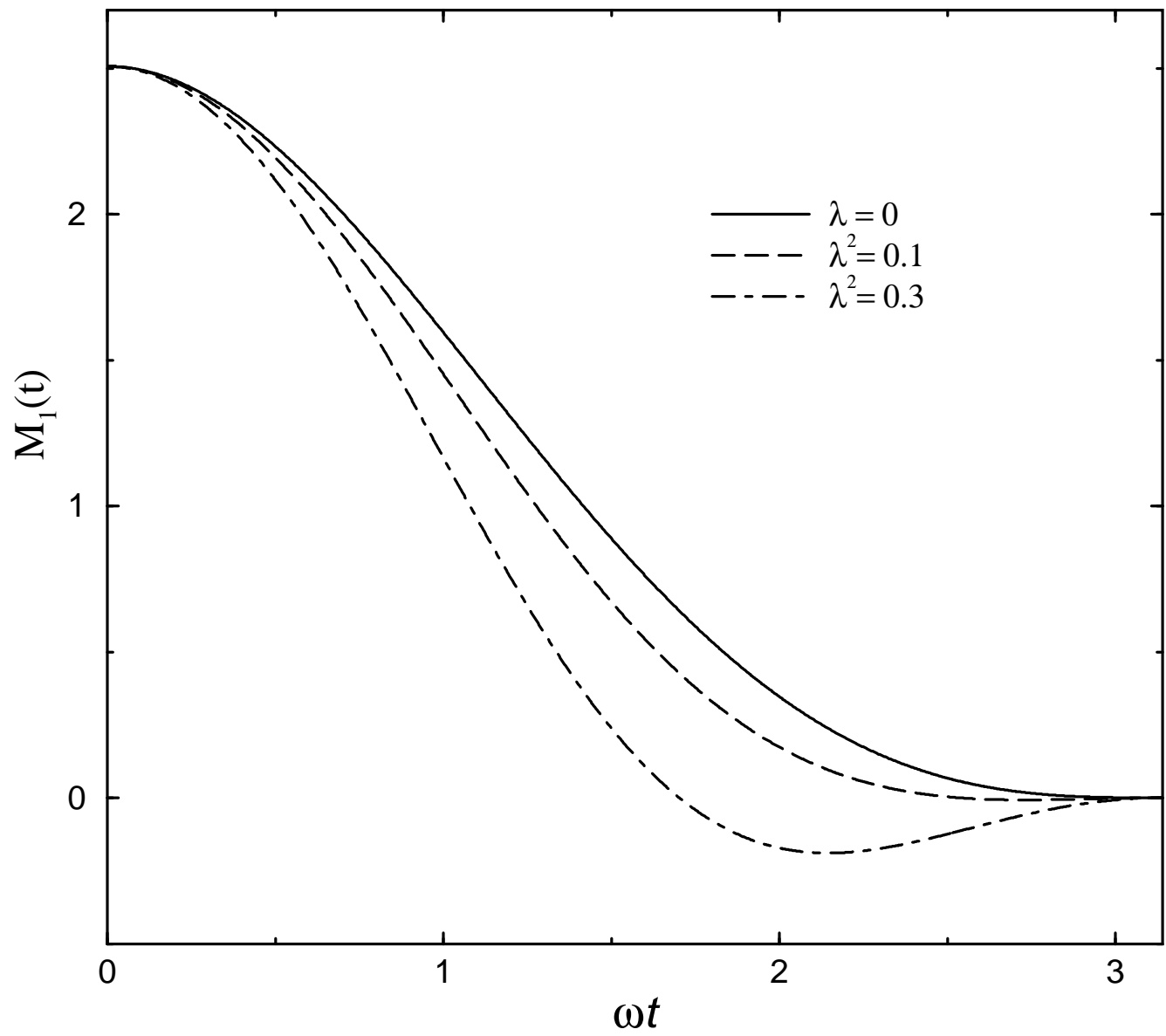
\end{document}